\documentclass[prd,twocolumn,showpacs,preprintnumbers,superscriptaddress,floatfix,nofootinbib]{revtex4-1}
\usepackage{amsfonts}
\usepackage{graphicx,,booktabs}
\usepackage{amssymb}
\usepackage{amsmath,txfonts}
\usepackage{graphicx}
\usepackage{dcolumn}
\usepackage{color}
\usepackage{bm}
\usepackage[colorlinks,
            citecolor=blue,
            anchorcolor=green,
            menucolor=orange,
            linkcolor=red,
            filecolor=red,
            runcolor=pink,
            urlcolor=blue,
            frenchlinks=red]{hyperref}

\begin{document}

\title{ Pion-induced production of hidden-charm pentaquarks $%
P_{c}(4312),P_{c}(4440)$, and $P_{c}(4457)$}
\author{Xiao-Yun Wang}
\thanks{xywang01@outlook.com}
\affiliation{Department of physics, Lanzhou University of Technology,
Lanzhou 730050, China}
\author{Jun He}
\thanks{Corresponding author : junhe@njnu.edu.cn}
\affiliation{Department of Physics and Institute of Theoretical Physics, Nanjing Normal University,
Nanjing, Jiangsu 210097, China}
\author{Xu-Rong Chen}
\affiliation{Institute of Modern Physics, Chinese Academy of Sciences, Lanzhou 730000, China}
\author{Quanjin Wang}
\affiliation{Lanzhou University of Technology,
Lanzhou 730050, China}
\author{Xinmei Zhu}
\affiliation{Department of Physics, Yangzhou University, Yangzhou, 225009, , People's Republic of China}

\begin{abstract}
The production of the hidden-charm pentaquarks $P_{c}$ via pion-induced reaction on a
proton target is investigated within an effective Lagrangian approach. Three experimentally observed states, $P_c(4312)$, $P_c(4440)$, and $P_c(4457)$, are considered in the calculation, and the
Reggeized $t$-channel meson exchange is considered as  main background
for the reaction $\pi ^{-}p\rightarrow J/\psi n$. The numerical results show
that the experimental data of the total cross section of the reaction $\pi
^{-}p\rightarrow J/\psi n$  at $W\simeq 5$ GeV can be well explained
by contribution of the Reggeized $t$ channel with reasonable cutoff. If the
branching ratios $Br[P_{c}\rightarrow J/\psi N]\simeq 3\%$ and $%
Br[P_{c}\rightarrow \pi N]\simeq 0.05\%$ are taken, the average value
of the cross section from the $P_{c}(4312)$ contribution is about 1.2 nb/100
MeV, which is consistent with existing rude data at near-threshold energies.
The  results indicate that the branching ratios of the $P_{c}$ states
 to the $J/\psi N$ and $\pi N$ should be small. The shape of differential
cross sections shows that the Reggeized $t$-channel provides a sharp increase
at extreme forward angles, while the differential cross sections from the $P_{c}$ states
contributions are relatively flat. High-precision experimental measurements
on the reaction $\pi ^{-}p\rightarrow J/\psi n$   at near-threshold
energies are suggested to confirm the LHCb hidden-charm pentaquarks as genuine states, and such experiments are also helpful to understand the origin of  these resonance structures.
\end{abstract}

\maketitle

\section{Introduction}

Study of the exotic hadrons beyond the constituent quark model is an important way to understand how quarks combine to form a hadron. As of now, many hidden-charm exotic states have been
observed and listed in the Review of Particle Physics (PDG)~\cite%
{Tanabashi:2018oca}.   Different from the charmonium-like states, only three possible candidates for the hidden-charm pentaquarks were reported in the literature~\cite{Aaij:2019vzc}. The studies of the hidden-charm pentaquarks in both experiment and theory are very important to understand the exotic hadrons.

The observation of two possible candidates of hidden-charm pentaquarks,  $P_{c}(4450)$ and $P_{c}(4380)$, was reported by the LHCb Collaboration in 2015 \cite{Aaij:2015tga}, which is an important progress of the search for the exotic hadrons. Recently, LHCb updated their results of the hidden-charm pentaquark~\cite{Aaij:2019vzc}.  The $P_c(4450)$ splits into two narrower structure with masses of  $4440.3\pm 1.3_{-4.7}^{+4.1}\mbox{ MeV}$ and $4457.3\pm 0.6_{-1.7}^{+4.1}\mbox{ MeV}$, which are renamed as $P_c(4440)$ and $P_c(4457)$, respectively.  An additional state $P_c(4312)$ with a mass of $4311.9\pm 0.7_{-0.6}^{+6.8}\mbox{ MeV}$  was also reported. After the result was released, the
theoretical researches on these three $P_{c}$ states is springing up~\cite
{Chen:2019asm,Chen:2019bip,Liu:2019tjn,He:2019ify,Huang:2019jlf,Ali:2019npk,Xiao:2019mvs,Shimizu:2019ptd,Guo:2019kdc,Xiao:2019aya,Guo:2019fdo,Cao:2019kst,Mutuk:2019snd,Weng:2019ynv,Zhu:2019iwm,Zhang:2019xtu,Eides:2019tgv,Wang:2019got,Meng:2019ilv,Cheng:2019obk}.

In fact, after the observation in 2015, many theoretical interpretations of  these structures were proposed~\cite{Chen:2015loa,Chen:2015moa,Roca:2015dva,Guo:2015umn,Maiani:2015vwa,He:2015cea,Liu:2015fea}.  The  $P_c(4450)$ and $P_c(4380)$ are close to the $\Sigma_c\bar{D}^*$ and $\Sigma_c^*\bar{D}$ thresholds. It is very natural to assign them as two molecular states of $\Sigma_c\bar{D}^*$ and $\Sigma_c^*\bar{D}$, respectively~\cite{He:2015cea,He:2016pfa}.  However, the opposite parities of  these two states make it difficult to assign both states as S-wave molecular states.  The new LHCb results shows that the previous $P_c(4450)$ structure  should be composed of two peaks of the $P_c(4440)$ and $P_c(4457)$.  It is reasonable to expect that puzzling spin parities are from the low precision of previous observation and will be changed with more data accumulated.  Combined with the observation of the $P_c(4312)$ which is close to the $\Sigma_c\bar{D}$ threshold, these three states can be naturally interpreted as the S-wave molecular states.   There are only two S-wave $\Sigma_c\bar{D}^*$ state with spin parities $1/2^-$ and $3/2^-$, and only  one S-wave $\Sigma_c\bar{D}$ state with $1/2^-$.  In fact, before the LHCb observation of the $P_c$ states, there have been many predictions of the hidden-charm pentaquarks~\cite{Wang:2011rga,Yang:2011wz,Wu:2012md}, the calculations in which support the existence of such bound states.  It is interesting to see that two states $P_c(4440)$ and $P_c(4457)$ were observed near $\Sigma_c\bar{D}^*$ threshold and one state $P_c(4312)$ was observed near $\Sigma_c\bar{D}$ threshold.
It was further supported by the theoretical studies after the new LHCb results released~\cite{Chen:2019asm,Liu:2019tjn,He:2019ify,Huang:2019jlf,Xiao:2019mvs,Xiao:2019aya}%
, that is, the $P_{c}(4312)$ can be assigned as a S-wave $\Sigma _{c}\bar{D}$ bound
state with spin parity $1/2^{-}$, and the $P_{c}(4440)$ and $P_{c}(4457)$ as
S-wave $\Sigma _{c}\bar{D}^{\ast }$ bound states with spin parities $1/2^{-}$
and $3/2^{-}$, respectively.

Although the molecular-state interpretation is quite consistent with the current LHCb observation, there are still many other proposals to understand the origin of these $P_c$ states~\cite
{Chen:2019asm,Chen:2019bip,Liu:2019tjn,He:2019ify,Huang:2019jlf,Ali:2019npk,Xiao:2019mvs,Shimizu:2019ptd,Guo:2019kdc,Xiao:2019aya,Guo:2019fdo,Cao:2019kst,Mutuk:2019snd,Weng:2019ynv,Zhu:2019iwm,Zhang:2019xtu,Eides:2019tgv,Wang:2019got,Meng:2019ilv,Cheng:2019obk}.  Moreover, up to now, these hidden-charm pentaquarks were still only observed in the $\Lambda _{b}$
decay at LHCb. If these states is really composed of a charm quark pair and three light quark, it should be easy to be produced by striking a nucleon by a particle, such as nucleon,  photon or pion,  to excite a charm quark pair in the nucleon. Confirmation of the existence of these resonance structure in the photon-  or pion-induced production is very important to establish the $P_c$ states as  genuine particles.  Hence, it is urgent to search for the hidden-charm pentaquark in such production processes.

The productions of the hidden-charm pentaquarks with nucleon target are proposed even before the LHCb's first observation of the  hidden-charm pentaquarks $P_{c}(4450)$ and $P_{c}(4380)$.  In  Ref.~\cite{Wu:2010jy}
the hidden-charm pentaquark states were predicted and suggested to be looked for in the reaction $p\bar{p}\rightarrow p\bar{p}J/\psi $. In Ref.~\cite{Huang:2013mua}, the photoproduction of the hidden-charm pentaquark was
first suggested to be applied at Jefferson Laboratory.  Then,
the productions of these predicted pentaquark state via pion- or kaon-induced
reaction were calculated \cite{Wang:2015qia,Wang:2015xwa}.  After the states $P_{c}(4450)$ and $%
P_{c}(4380)$ were experimentally observed, many theoretical calculations
about these two pentaquarks produced in different processes
appeared~\cite%
{Kim:2016cxr,Wang:2015jsa,Lu:2015fva,Garzon:2015zva,Karliner:2015voa,Meziani:2016lhg}.  After new LHCb results was released, we also updated the predictions about the photoproductions of  three $P_c$ states~\cite{Wang:2019krd}, which is accessible at JLab.

\begin{figure}[tbph]
\begin{center}
\includegraphics[scale=0.52]{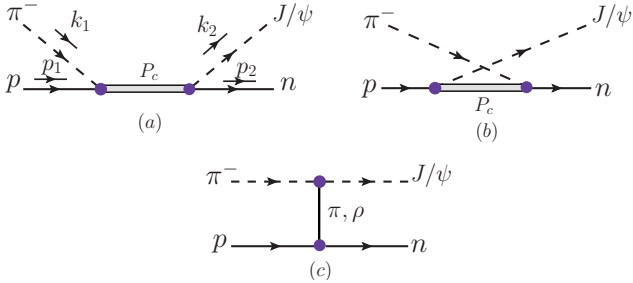}
\end{center}
\caption{(Color online) Feynman diagrams for the $\protect\pi %
^{-}p\rightarrow J/\protect\psi n$ reaction.}
\label{Fig: Feynman}
\end{figure}

Another important way to study the hidden-charm pentaquarks is pion-induced production~\cite{Wang:2015qia,Kim:2016cxr}.
At present, there are some experimental data for the $\pi ^{-}p\rightarrow
J/\psi n$ reaction \cite{Jenkins:1977xb,Chiang:1986gn}. Furthermore, we note
that $P_{c}(4312),P_{c}(4440)$ and $P_{c}(4457)$ are all observed on the $%
J/\psi p$ invariant mass spectrum. In terms of experiments, the secondary pion beam is accessible at J-PARC~%
\cite{Austregesilo:2018mno,Kumano:2015gna} and COMPASS~\cite{Nerling:2012er}
with high intensity.  Therefore, combining these experimental
data to examine the role of these three states in pion--induced reaction is
very necessary. The reaction mechanisms of the reaction $\pi ^{-}p\rightarrow J/\psi n$ are illustrated
in Fig.~\ref{Fig: Feynman}. These include the production of pentaquark $%
P_{c} $ states via $s$- and $u$-channel (as shown in Fig.~\ref{Fig: Feynman}
$(a)-(b)$), and $t$-channel $\pi $ and $\rho $ exchanges as depicted in Fig.~%
\ref{Fig: Feynman} $(c)$. Considering the off-shell
effect of the intermediate $P_{c}$ states, the contribution from $u$-channel
will be omitted.

In this work, within the frame of an effective Lagrangian approach, the productions of  $P_{c}$  states via
pion-induced reaction on a proton target will be investigated. In the calculation, three hidden-charm pentaquarks, $P_c(4457)$, $P_c(4440)$, and $P_c(4312)$,  which are assumed to carry spin parities $3/2^-$, $1/2^-$, and $1/2^-$, respectively, will be considered in the calculation. The Reggeized treatment will be applied to $t$ channel to describe the main background of the productions.

This paper is organized as follows. After the Introduction, we present the
formalism including Lagrangians and amplitudes of the $P_{c}$ states
productions in Section II. The numerical results of the cross section follow
in Section III. Finally, the paper{\ ends} with a brief summary.

\section{Formalism}

\subsection{Lagrangians}

To gauge the contributions of Fig.~\ref{Fig: Feynman}, one needs the
following Lagrangians for the $s$-channel $P_{c}$ exchanges \cite%
{Kim:2011rm,Oh:2007jd,Zou:2002yy,Wang:2015qia,Wang:2015xwa},
\begin{eqnarray}
\mathcal{L}_{\pi NP_{c}}^{1/2^{-}} &=&g_{_{\pi NP_{c}}}^{1/2^{-}}\bar{N}\vec{%
\tau}\cdot \vec{\pi}P_{c}+{\rm H.c.}, \\
\mathcal{L}_{P_{c}\psi N}^{1/2^{-}} &=&g_{P_{c}\psi N}^{1/2^{-}}\bar{N}%
\gamma _{5}\gamma _{\mu }P_{c}\psi ^{\mu }+{\rm H.c.}, \\
\mathcal{L}_{\pi NP_{c}}^{3/2^{-}} &=&\frac{g_{_{\pi NP_{c}}}^{3/2^{-}}}{%
m_{\pi }}\bar{N}\gamma _{5}\vec{\tau}\cdot \partial _{\mu }\vec{\pi}%
P_{c}^{\mu }+{\rm H.c.}, \\
\mathcal{L}_{P_{c}\psi N}^{3/2^{-}} &=&\frac{-ig_{P_{c}\psi N}^{3/2^{-}}}{%
2m_{N}}\bar{N}\gamma _{\nu }\psi ^{\mu \nu }P_{c\mu }-\frac{g_{2}}{%
(2m_{N})^{2}}\partial _{\nu }\bar{N}\psi ^{\mu \nu }P_{c\mu }  \notag \\
&&+\frac{g_{2}}{(2m_{N})^{2}}\bar{N}\partial _{\nu }\psi ^{\mu \nu }P_{c\mu
}+{\rm H.c.},  \label{L2}
\end{eqnarray}%
where ${N}$, ${\pi}$, $P_{c}$ and $\psi $ are the nucleon, the pion, the $P_{c}$
state and the $J/\psi $ meson fields, respectively, and ${\tau}$ is the Pauli
matrix.

The values of $g_{_{\pi NP_{c}}}^{1/2^{-}}$ and $g_{_{\pi NP_{c}}}^{3/2^{-}}$
can be determined by the corresponding decay widths,%
\begin{eqnarray}
\Gamma _{P_{c}\rightarrow \pi N}^{1/2^{-}} &=&\frac{3g_{\pi
NP_{c}}^{2}(m_{N}+E_{N})}{4\pi m_{P_{c}}}|\vec{p}_{N}^{~\mathrm{c.m.}}|,\\
\Gamma _{P_{c}\rightarrow \pi N}^{3/2^{-}} &=&\frac{g_{\pi
NP_{c}}^{2}(m_{N}-E_{N})}{4\pi m_{\pi }^{2}m_{P_{c}}}|\vec{p}_{N}^{~\mathrm{%
c.m.}}|^{3},
\end{eqnarray}%
with%
\begin{eqnarray}
|\vec{p}_{N}^{~\mathrm{c.m.}}| &=&\frac{\lambda (m_{P_{c}}^{2},m_{\pi
}^{2},m_{N}^{2})}{2m_{P_{c}}}, \\
E_{N} &=&\sqrt{|\vec{p}_{N}^{~\mathrm{c.m.}}|^{2}+m_{N}^{2}},
\end{eqnarray}%
where $\lambda $ is the K$\ddot{a}$llen function with a definition of $%
\lambda (x,y,z)=\sqrt{(x-y-z)^{2}-4yz}$. For the values of $g_{P_{c}\psi
N}^{1/2^{-}}$ and $g_{P_{c}\psi N}^{3/2^{-}}$, we have conducted relevant
research discussions in our previous studies about the photoproduction of the $P_c$ states~\cite{Wang:2019krd}. Obviously, the coupling
constants of the $g_{P_{c}\psi N}$ and $g_{\pi NP_{c}}$ are proportional to
the corresponding values of decay width of $P_{c}$ states. In Table.~\ref%
{tab:fit1}, we present the values of coupling constants by assuming the $%
J/\psi N$ and $\pi N$ channels account for 3\% and 0.05\% of total widths of
the $P_{c}$ states, respectively.

\renewcommand\tabcolsep{0.5cm} \renewcommand{\arraystretch}{1.4}
\begin{table}[h]
\caption{The values of coupling constants by assuming that the $J/\protect\psi N$
and $\protect\pi N$ channels account for 3\% and 0.05\% of total widths of
the $P_{c}$ states, respectively.}
\label{tab:fit1}%
\begin{tabular}{lccc}
\toprule[1.5pt] & $P_{c}(4312)$ & $P_{c}(4440)$ & $P_{c}(4457)$ \\ \hline
$g_{P_{c}\psi N}$ & $0.06$ & $0.08$ & $0.036$ \\
$g_{_{\pi NP_{c}}}$ & $0.0036$ & $0.0053$ & $0.0005$ \\
\bottomrule[1.5pt] &  &  &
\end{tabular}%
\end{table}

For the $t$-channel via $\pi $ and $\rho $ exchanges, the effective
Lagrangians read as%
\begin{eqnarray}
\mathcal{L}_{\psi \pi \pi } &=&-ig_{_{\psi \pi \pi }}(\pi ^{-}\partial _{\mu
}\pi ^{+}-\partial _{\mu }\pi ^{-}\pi ^{+})\psi ^{\mu }, \\
\mathcal{L}_{\psi \rho \pi } &=&\frac{g_{_{\psi \rho \pi }}}{m_{\psi }}%
\epsilon ^{\mu \nu \alpha \beta }\partial _{\mu }\psi _{\nu }\partial
_{\alpha }\rho _{\beta }\cdot \pi , \\
\mathcal{L}_{\pi NN} &=&-ig_{_{\pi NN}}\bar{N}\gamma _{5}\vec{\tau}\cdot
\vec{\pi}N, \\
\mathcal{L}_{\rho NN} &=&-g_{_{\rho NN}}\bar{N}[\gamma _{\mu }-\frac{\kappa
_{\rho NN}}{2m_{N}}\sigma _{\mu \nu }\partial ^{\nu }]\vec{\tau}\cdot \rho
^{\mu }N,
\end{eqnarray}%
where ${N}$, $\pi $, $\rho $ and $\psi $ are the nucleon, the pion, the $\rho $ and the $%
J/\psi $ meson fields, respectively. Here, the $g_{_{\pi NN}}^{2}/4\pi
=12.96,g_{_{\rho NN}}=3.36,$ and $\kappa _{\rho NN}=6.1$ are adopted \cite%
{Lin:1999ve,Baru:2011bw,Lu:2015fva}. Moreover, The $g_{_{\psi \pi \pi
}}=8.2\times 10^{-4}$ and $g_{_{\psi \rho \pi }}=3.2\times 10^{-2}$ will be
used in the calculations, which was mentioned in  Refs.~\cite%
{Wu:2013xma,Lu:2015fva}.

\subsection{Amplitudes}

According to  above Lagrangians, the scattering amplitude of the reaction $\pi
^{-}p\rightarrow J/\psi n$  can be written as%
\begin{equation}
-i\mathcal{M}=\epsilon _{J/\psi }^{\mu }(k_{2})\bar{u}(p_{2})\mathcal{A}%
_{\mu }u(p_{1}),
\end{equation}%
where $\epsilon _{J/\psi }^{\mu }$ is the polarization vector of the $J/\psi $
meson, and $u$ is the Dirac spinor of the nucleon.

The reduced amplitudes $\mathcal{A}_{i,\mu }$ for the $s$ channel with each $%
J^{P}$ assignment of $P_{c}$ state and the $t$-channel are written as
\begin{eqnarray}
\mathcal{A}_{s}^{P_{c}(1/2^{-})} &=&\sqrt{2}g_{_{\pi
NP_{c}}}^{1/2^{-}}g_{P_{c}\psi N}^{1/2^{-}}\mathcal{F}_{s}(q_{s}^{2})\gamma
_{5}\gamma _{\mu }\frac{(\rlap{$\slash$}q+m_{P_{c}})}{%
s-m_{P_{c}}^{2}+im_{P_{c}}\Gamma _{P_{c}}}  \label{AmpT1} \\
\mathcal{A}_{s}^{P_{c}(3/2^{-})} &=&\sqrt{2}\frac{g_{_{\pi NP_{c}}}^{3/2^{-}}%
}{m_{\pi }}\frac{-ig_{P_{c}\psi N}^{3/2^{-}}}{2m_{N}}\mathcal{F}%
_{s}(q_{s}^{2})\gamma _{\sigma }(k_{2}^{\beta }g^{\mu \sigma }-k_{2}^{\sigma
}g^{\mu \beta })  \notag \\
&&\times \frac{(\rlap{$\slash$}q+m_{P_{c}})}{s-m_{P_{c}}^{2}+im_{P_{c}}%
\Gamma _{P_{c}}}\Delta _{\beta \alpha }k_{1}^{\alpha }\gamma _{5}, \\
\mathcal{A}_{t}^{\pi } &=&\sqrt{2}g_{_{\psi \pi \pi }}g_{_{\pi NN}}\frac{%
\mathcal{F}_{t}(q_{t}^{2})}{t-m_{\pi }^{2}}\gamma _{5}k_{1\mu }, \\
\mathcal{A}_{t}^{\rho } &=&\sqrt{2}\frac{g_{_{\psi \rho \pi }}}{m_{\psi }}%
g_{_{\rho NN}}\mathcal{F}_{t}(q_{t}^{2})\frac{\mathcal{P}^{\nu \xi }}{%
t-m_{\rho }^{2}}\epsilon ^{\mu \nu \alpha \beta }k_{2\alpha
}(k_{2}-k_{1})_{\beta }  \notag \\
&&[\gamma _{\xi }+\frac{\kappa _{\rho NN}}{4m_{N}}(\gamma _{\xi }%
\rlap{$\slash$}q_{t}-\rlap{$\slash$}q_{t}\gamma _{\xi })],
\end{eqnarray}%
with
\begin{eqnarray}
\Delta _{\beta \alpha } &=&-g_{\beta \alpha }+\frac{1}{3}\gamma ^{\beta
}\gamma ^{\alpha }  \notag \\
&&+\frac{1}{3m_{P_{c}}}(\gamma ^{\beta }q^{\alpha }-\gamma ^{\alpha
}q^{\beta })+\frac{2}{3m_{P_{c}}^{2}}q^{\beta }q^{\alpha }, \\
\mathcal{P}^{\nu \xi } &=&i\left( g^{\nu \xi }+q_{\rho }^{\nu }q_{\rho
}^{\xi }/m_{\rho }^{2}\right) ,
\end{eqnarray}%
where $s=(k_{1}+k_{2})^{2}$ and $t=(k_{1}-k_{2})^{2}$ is the Mandelstam
variables. For the $s$-channel $P_{c}$-state exchange, a general form factor
is adopted to describe the size of hadrons, i.e. \cite%
{Wang:2015jsa,Wang:2017qcw},

\begin{equation}
\mathcal{F}_{s}(q_{s}^{2})=\frac{\Lambda _{s}^{4}}{\Lambda
_{s}^{4}+(q_{s}^{2}-m_{P_{c}}^{2})^{2}}~,
\end{equation}%
where $q_{s}$ and $m_{P_{c}}$ are 4-momentum and mass of the exchanged $%
P_{c}$ state, respectively. Considering that it is a{\ heavier hadron
production, the }typical value of cut off $\Lambda _{s}=0.5$ GeV will be{\
taken }as used in Refs. \cite{Kim:2011rm,Wang:2015jsa}.

For the $t$-channel meson exchanges \cite%
{Wang:2015xwa,Liu:2008qx,Lu:2015fva,Wan:2015gsl,Wang:2017qcw,Wang:2017plf,Wang:2015hfm,Wang:2018mjz}%
, the general form factor $\mathcal{F}_{t}(q_{t}^{2})$ consisting of $%
\mathcal{F}_{\psi V\pi }=(\Lambda _{t}^{2}-m_{V}^{2})/(\Lambda
_{t}^{2}-q_{V}^{2})$ and $\mathcal{F}_{VNN}=(\Lambda
_{t}^{2}-m_{V}^{2})/(\Lambda _{t}^{2}-q_{V}^{2})$ are taken into account.
Here, $q_{V}$ and $m_{V}$ are 4-momentum and mass of the exchanged meson,
respectively. The value of the cutoff $\Lambda _{t}$ will be discussed in
the next section.

\subsection{Reggeized $t$-channel}

\label{Sec: Regge}

The Reggeized treatment is often adopted to analyze hadron production at high energies\cite
{Wan:2015gsl,Wang:2015xwa,Haberzettl:2015exa,Wang:2015hfm,Ozaki:2009wp,Wang:2018mjz,Wang:2017qcw,Wang:2017plf}. It can be introduced by replacing the $t$-channel Feynman
propagator by the Regge propagator as,
\begin{eqnarray}
\frac{1}{t-m_{\pi }^{2}} &\rightarrow &(\frac{s}{s_{scale}})^{\alpha _{\pi
}(t)}\frac{\pi \alpha _{\pi }^{\prime }}{\Gamma \lbrack 1+\alpha _{\pi
}(t)]\sin [\pi \alpha _{\pi }(t)]}, \\
\frac{1}{t-m_{\rho}^{2}} &\rightarrow &(\frac{s}{s_{scale}})^{\alpha
_{\rho }(t)-1}\frac{\pi \alpha _{\rho }^{\prime }}{\Gamma \lbrack \alpha
_{\rho }(t)]\sin [\pi \alpha _{\rho }(t)]},
\end{eqnarray}%
where the scale factor $s_{scale}$ is fixed at 1 GeV. Moreover, the Regge
trajectories of $\alpha _{\pi }(t)$ and $\alpha _{\rho }(t)$ read as \cite%
{Liu:2008qx,Kim:2016cxr,Wang:2017plf},%
\begin{equation}
\alpha _{\pi }(t)=0.7(t-m_{\pi }^{2}),\ \alpha _{\rho }(t)=0.55+0.8t.\quad \
\
\end{equation}%
One can observe that  no additional parameter is
introduced after the Reggeized treatment is introduced,.

\section{Numerical results}

After  above preparation, the cross section of the reaction $\pi ^{-}p\rightarrow
J/\psi n$  can be calculated and compared with experimental data
\cite{Jenkins:1977xb,Chiang:1986gn}. The differential cross section in the
center of mass (c.m.) frame is written as
\begin{equation}
\frac{d\sigma }{d\cos \theta }=\frac{1}{32\pi s}\frac{\left\vert \vec{k}%
_{2}^{{~\mathrm{c.m.}}}\right\vert }{\left\vert \vec{k}_{1}^{{~\mathrm{c.m.}}%
}\right\vert }\left( \frac{1}{2}\sum\limits_{\lambda }\left\vert \mathcal{M}%
\right\vert ^{2}\right) ,
\end{equation}%
where $s=(k_{1}+p_{1})^{2}$, and $\theta $ denotes the angle of the outgoing
$J/\psi $ meson relative to the $\pi $ beam direction in the c.m. frame. $\vec{k}%
_{1}^{{~\mathrm{c.m.}}}$ and $\vec{k}_{2}^{{~\mathrm{c.m.}}}$ are the
three-momenta of the initial $\pi $ beam and final $J/\psi $, respectively.

\begin{figure}[h!]
\begin{center}
\includegraphics[scale=0.4]{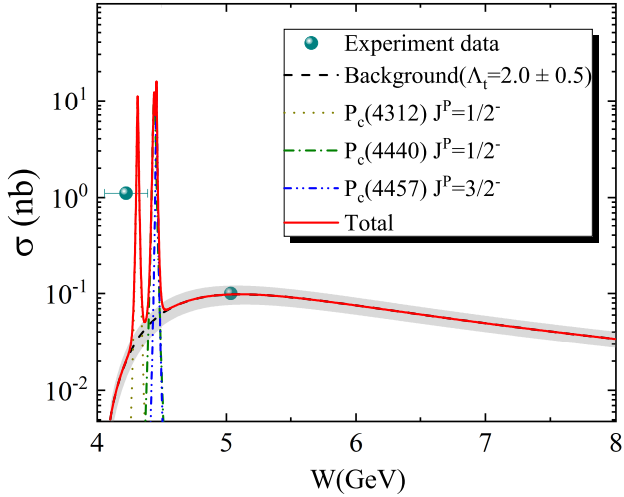}
\end{center}
\caption{(Color online) Total cross section for the  reaction $\protect\pi %
^{-}p\rightarrow J/\protect\psi n$. The black dashed, dark yellow
dotted, green dot-dashed, blue dash-double-dotted, and red solid lines are
for the background, the $P_{c}(4312)$, the $P_{c}(4440),$ the $P_{c}(4457)$ and total
contributions, respectively. The bands stand for the error bar of the cutoff
$\Lambda _{t}$. The experimental data are from Refs.\protect\cite%
{Jenkins:1977xb,Chiang:1986gn}.}
\label{Fig:total01}
\end{figure}

\begin{figure}[h!]
\centering
\includegraphics[scale=0.4]{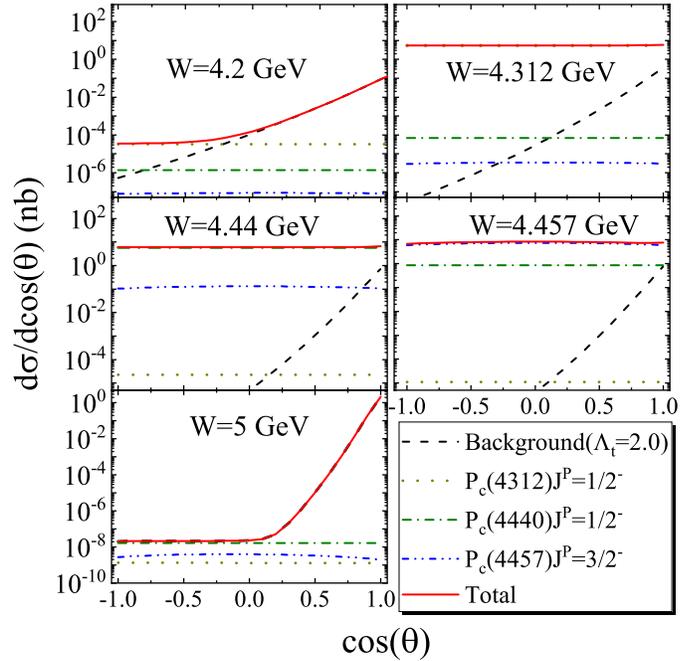}
\caption{(Color online) The differential cross section $d\protect\sigma %
/d\cos \protect\theta $ of the $\protect\pi ^{-}p\rightarrow J/\protect\psi %
n $ process as a function of $\cos \protect\theta $ at different c.m.
energies. The black dashed, dark yellow dotted, green dot-dashed, blue
dash-double-dotted, and red solid lines are for the background, $%
P_{c}(4312),P_{c}(4440),$ $P_{c}(4457)$ and total contributions,
respectively.}
\label{dcs01}
\end{figure}

In this work,  total and differential cross sections of the  reaction $\pi
^{-}p\rightarrow J/\psi n$ are calculated as presented in Figs. \ref%
{Fig:total01}-\ref{dcs01}. For the total cross section of the reaction $\pi
^{-}p\rightarrow J/\psi n$, there are currently two experimental
data points located near the energy threshold and at center of mass (c.m.)
energy $W\simeq 5$ GeV, respectively. We found that it is difficult to meet both
experimental data points if considering only the $t$-channel background
contribution. It can be seen from Fig. \ref{Fig:total01} that the data
points at $W\simeq 5$ GeV are well matched by the cross section of $t$%
-channel by taking a cutoff  $\Lambda _{t}=2$ GeV. However,  at the same time, the data
point near the threshold is more than an order of magnitude larger than the theoretical
value of $t$-channel contribution. If we consider
the contribution from the $s$-channel $P_{c}$ state, the data point near the
threshold can be well explained. As shown in Fig.~\ref{Fig:total01}, one
find that  experimental data point near the threshold  is consistent with
the contribution from the $P_{c}(4312)$ state by assuming  branching
ratios $Br[P_{c}\rightarrow J/\psi N]\simeq 3\%$ and $Br[P_{c}\rightarrow \pi
N]\simeq 0.05\%$. Due to adoption of the Regge propagator, we find that the
cross section of the $t$ channel reaches a maximum at $W\simeq 5$ GeV, and
 the total cross section decreases as the energy increases. If the Feynman
propagator is adopted, the  total cross section from  $t$ channel
would become larger and larger with the increase of the c.m. energy. The
difference between the Regge model and the Feynman model will help to
clarify the role of Regge propagator in the future experiment.

\begin{figure}[h!]
\begin{center}
\includegraphics[scale=0.4]{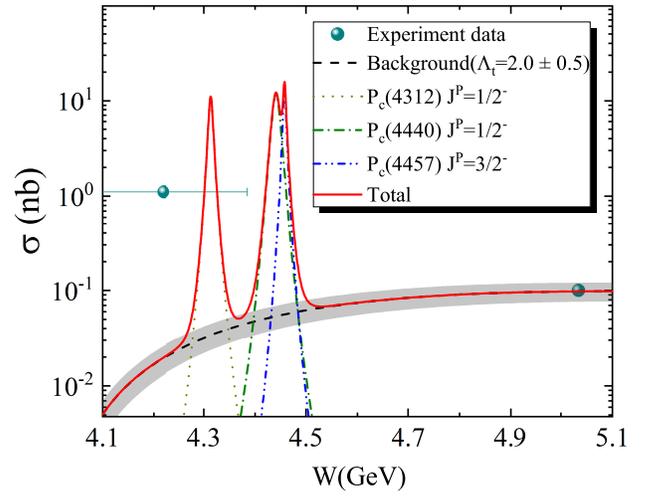}
\end{center}
\caption{(Color online) Same as Fig. 2 except that the energy range is
reduced.}
\label{Fig:total02}
\end{figure}

In order to distinguish the contributions from the three $P_{c}$
states more clearly, the Fig.~\ref{Fig:total02} is presented, which is the
same as the Fig.~\ref{Fig:total01} except that the energy range is reduced.
From Fig.~\ref{Fig:total02}, one can see three distinct peaks, which are
 from the contributions of three $P_{c}$ states. As we discussed in
our previous work about the photoproduction of the $P_{c}$ states\cite{Wang:2019krd}, the $P_{c}(4312)$ can be observed
within a bin of 0.1 GeV. But if one wants to distinguish two peaks from the  $%
P_{c}(4440) $ and $P_{c}(4457)$, a bin at least at an order of 10 MeV is
required. According to our calculation by assuming the branching ratios $%
Br[P_{c}\rightarrow J/\psi N]\simeq 3\%$ and $Br[P_{c}\rightarrow \pi
N]\simeq 0.05\%$, if the width of a bin is 0.1GeV, the theoretical average value
of the cross section from the $P_{c}(4312)$ contribution is about 1.2 nb in
a bin interval, which is just in agreement with the experimental value near
the threshold.

In Fig. \ref{dcs01}, we present our prediction of the differential cross
section of the reaction  $\pi ^{-}p\rightarrow J/\psi n$  at different c.m.
energy. It can be seen that the differential cross section has a large
contribution at forward angles, which is caused by the Reggeized $t$%
-channel. In addition, we find that the shape of differential cross section
tends to be flat at the c.m. energy $W=4.312$, $4.44$ and 4.457 GeV, which
is due to the large contributions of the $P_{c}$ state at these energy
points. Since the spin-parity quantum numbers of these $P_{c}$ states are
selected to be $1/2^{-}$ or $3/2^{-}$, these $P_{c}$ states can couple to
both the initial $\pi N$ and final $J/\psi N$ in S wave, and the couplings by
higher partial waves can be ignored because the momentum between the final $%
J/\psi N$ is very small. Therefore, the shape of the differential cross section of
these $P_{c}$ states is relatively flat, which reflects the characteristics of the
S-wave coupling.

\section{Summary and discussion}

We have studied the reaction  $\pi ^{-}p\rightarrow J/\psi n$ within the Regge model.
The numerical results show that the experimental data near the threshold
 can be well explained if we consider the contribution from the
pentaquark states. In addition, numerical results also indicate the
experimental data at $W\simeq 5$ GeV is unlikely to come from the
contribution of the $P_c$ state but should be caused by the $t$-channel
background contribution.

At present, the branching ratios of $P_{c}$ decay to $J/\psi N$ and $\pi N$
are still undetermined, but if the branching ratios $Br[P_{c}\rightarrow
J/\psi N]\simeq 3\%$ and $Br[P_{c}\rightarrow \pi N]\simeq 0.05\%$ are
taken, then the average value of the cross section from the $P_{c}(4312)$
contribution is about 1.2 nb/100 MeV, which coincides with the experimental
data point near the threshold. Combined with the results in our previous
article about the photoproduction of the $P_c$ states~\cite{Wang:2019krd}, it is reasonable to think that the branching ratios of $%
P_{c}$ states to $J/\psi N$ and $\pi N$ should be relatively small.
Therefore, we suggest that experiments with high precision near the threshold
can be performed, which is very important for determining the
branch ratios and the internal structure of the $P_{c}$ states.

The differential cross sections for  the reaction $\pi ^{-}p\rightarrow J/\psi n$
are also calculated. One notices that the Reggeized $t$ channel is very
sensitive to the $\theta $ angle and gives considerable contributions at
forward angles. On the contrary, the shape of cross section from the $P_{c}$
states contribution is relatively flat, and it is related to the
spin-parity quantum numbers of these $P_{c}$ states, which can be checked by
future experiment and may be an effective way to examine the validity of the
Reggeized treatment and spin parities of these $P_{c}$ states.

J-PARC and COMPASS can generate pion beam
covering the above energy regions, and provide high-precision experimental
data. Our theoretical results will provide valuable reference information
for the studies of the pentaquark states at these facilities.

\section{Acknowledgments}

This project is supported by the National Natural Science Foundation of
China under Grants No. 11705076 and No. 11675228. We acknowledge the Natural
Science Foundation of Gansu province under Grant No. 17JR5RA113. This work
is partly supported by the HongLiu Support Funds for Excellent Youth Talents
of Lanzhou University of Technology.


\begin{thebibliography}{99}
\bibitem{Tanabashi:2018oca} M.~Tanabashi \textit{et al.}
[ParticleDataGroup], \textquotedblleft Review of Particle
Physics,\textquotedblright\ Phys.\ Rev.\ D \textbf{98}, 030001
(2018). 

\bibitem{Aaij:2019vzc} R.~Aaij \textit{et al.} [LHCb Collaboration],
\textquotedblleft Observation of a narrow pentaquark state, $P_{c}(4312)^{+}$%
, and of two-peak structure of the $P_{c}(4450)^{+}$,\textquotedblright\
arXiv:1904.03947 [hep-ex].

\bibitem{Aaij:2015tga} R.~Aaij \textit{et al.} [LHCb Collaboration],
``Observation of $J/\psi p$ Resonances Consistent with Pentaquark States in $%
\Lambda_b^0 \to J/\psi K^- p$ Decays,'' Phys.\ Rev.\ Lett.\ \textbf{115},
072001 (2015).


\bibitem{Chen:2019asm} R.~Chen, Z.~F.~Sun, X.~Liu and S.~L.~Zhu,
\textquotedblleft Strong LHCb evidence supporting the existence of the
hidden-charm molecular pentaquarks,\textquotedblright\ arXiv:1903.11013
[hep-ph].

\bibitem{Chen:2019bip} H.~X.~Chen, W.~Chen and S.~L.~Zhu, \textquotedblleft
Possible interpretations of the $P_{c}(4312)$, $P_{c}(4440)$, and $%
P_{c}(4457)$,\textquotedblright\ arXiv:1903.11001 [hep-ph].

\bibitem{Liu:2019tjn} M.~Z.~Liu, Y.~W.~Pan, F.~Z.~Peng, M.~S\'{a}nchez S\'{a}%
nchez, L.~S.~Geng, A.~Hosaka and M.~Pavon Valderrama, \textquotedblleft
Emergence of a complete heavy-quark spin symmetry multiplet: seven molecular
pentaquarks in light of the latest LHCb analysis,\textquotedblright\
arXiv:1903.11560 [hep-ph].

\bibitem{He:2019ify}
  J.~He,
  ``Study of $P_c(4457)$, $P_c(4440)$, and $P_c(4312)$ in a quasipotential Bethe-Salpeter equation approach,''
  Eur.\ Phys.\ J.\ C {\bf 79}, 393 (2019)

\bibitem{Huang:2019jlf} H.~Huang, J.~He and J.~Ping, \textquotedblleft
Looking for the hidden-charm pentaquark resonances in $J/\psi p$
scattering,\textquotedblright\ arXiv:1904.00221 [hep-ph].

\bibitem{Ali:2019npk}
  A.~Ali and A.~Y.~Parkhomenko,
  ``Interpretation of the narrow $J/\psi p$ Peaks in $\Lambda_b \to J/\psi p K^-$ decay in the compact diquark model,''
  Phys.\ Lett.\ B {\bf 793}, 365 (2019)

\bibitem{Xiao:2019mvs} C.~J.~Xiao, Y.~Huang, Y.~B.~Dong, L.~S.~Geng and
D.~Y.~Chen, \textquotedblleft Partial decay widths of $P_{c}(4312)$, $%
P_{c}(4440)$, and $P_{c}(4457)$ into $J/\psi p$ in a molecular
scenario,\textquotedblright\ arXiv:1904.00872 [hep-ph].

\bibitem{Shimizu:2019ptd} Y.~Shimizu, Y.~Yamaguchi and M.~Harada,
\textquotedblleft Heavy quark spin multiplet structure of $P_{c}(4312)$, $%
P_{c}(4440)$, and $P_{c}(4457)$,\textquotedblright\ arXiv:1904.00587
[hep-ph].

\bibitem{Guo:2019kdc}
  Z.~H.~Guo and J.~A.~Oller,
  ``Anatomy of the newly observed hidden-charm pentaquark states: $P_c(4312)$, $P_c(4440)$ and $P_c(4457)$,''
  Phys.\ Lett.\ B {\bf 793}, 144 (2019)

\bibitem{Xiao:2019aya} C.~W.~Xiao, J.~Nieves and E.~Oset, ``Heavy quark spin
symmetric molecular states from ${\bar D}^{(*)}\Sigma_c^{(*)}$ and other
coupled channels in the light of the recent LHCb pentaquarks,''
arXiv:1904.01296 [hep-ph].


\bibitem{Guo:2019fdo}
  F.~K.~Guo, H.~J.~Jing, U.~G.~Meißner and S.~Sakai,
  ``Isospin breaking decays as a diagnosis of the hadronic molecular structure of the $P_c(4457)$,''
  Phys.\ Rev.\ D {\bf 99}, 091501 (2019)



\bibitem{Cao:2019kst} X.~Cao and J.~p.~Dai, \textquotedblleft Pentaquark
photoproduction confronting with new LHCb observation,\textquotedblright\
arXiv:1904.06015 [hep-ph].

\bibitem{Mutuk:2019snd}
  H.~Mutuk,
  ``Neural Network Study of Hidden-Charm Pentaquark Resonances,''
  arXiv:1904.09756 [hep-ph].

\bibitem{Weng:2019ynv}
  X.~Z.~Weng, X.~L.~Chen, W.~Z.~Deng and S.~L.~Zhu,
  ``Hidden-charm pentaquarks and $P_c$ states,''
  arXiv:1904.09891 [hep-ph].


\bibitem{Zhu:2019iwm}
  R.~Zhu, X.~Liu, H.~Huang and C.~F.~Qiao,
  ``Implications of the hidden charm pentaquarks $P_c(X)$ at the LHCb,''
  arXiv:1904.10285 [hep-ph].

\bibitem{Zhang:2019xtu}
  J.~R.~Zhang,
  ``Exploring a $\Sigma_{c}\bar{D}$ state: with focus on $P_{c}(4312)^{+}$,''
  arXiv:1904.10711 [hep-ph].

\bibitem{Eides:2019tgv}
  M.~I.~Eides, V.~Y.~Petrov and M.~V.~Polyakov,
  ``New LHCb pentaquarks as hadrocharmonium states,''
  arXiv:1904.11616 [hep-ph].

\bibitem{Wang:2019got}
  Z.~G.~Wang,
  ``Analysis of the $P_c(4312)$, $P_c(4440)$, $P_c(4457)$ and related hidden-charm pentaquark states with QCD sum rules,''
  arXiv:1905.02892 [hep-ph].

\bibitem{Meng:2019ilv}
  L.~Meng, B.~Wang, G.~J.~Wang and S.~L.~Zhu,
  ``The hidden charm pentaquark states and $\Sigma_c\bar{D}^{(*)}$ interaction in chiral perturbation theory,''
  arXiv:1905.04113 [hep-ph].

\bibitem{Cheng:2019obk}
  J.~B.~Cheng and Y.~R.~Liu,
  ``$P_c(4457)^+$, $P_c(4440)^+$, and $P_c(4312)^+$: molecules or compact pentaquarks?,''
  arXiv:1905.08605 [hep-ph].


\bibitem{Chen:2015loa}
  R.~Chen, X.~Liu, X.~Q.~Li and S.~L.~Zhu,
  ``Identifying exotic hidden-charm pentaquarks,''
  Phys.\ Rev.\ Lett.\  {\bf 115}, 132002 (2015)


\bibitem{Chen:2015moa}
  H.~X.~Chen, W.~Chen, X.~Liu, T.~G.~Steele and S.~L.~Zhu,
  ``Towards exotic hidden-charm pentaquarks in QCD,''
  Phys.\ Rev.\ Lett.\  {\bf 115}, 172001 (2015)

\bibitem{Roca:2015dva}
  L.~Roca, J.~Nieves and E.~Oset,
  ``LHCb pentaquark as a $\bar{D}^*\Sigma_c-\bar{D}^*\Sigma_c^*$ molecular state,''
  Phys.\ Rev.\ D {\bf 92}, 094003 (2015)


\bibitem{Guo:2015umn}
  F.~K.~Guo, U.~G.~Meißner, W.~Wang and Z.~Yang,
  ``How to reveal the exotic nature of the P$_c$(4450),''
  Phys.\ Rev.\ D {\bf 92}, 071502 (2015)


\bibitem{Maiani:2015vwa}
  L.~Maiani, A.~D.~Polosa and V.~Riquer,
  ``The New Pentaquarks in the Diquark Model,''
  Phys.\ Lett.\ B {\bf 749}, 289 (2015)


\bibitem{He:2015cea}
  J.~He,
  ``$\bar{D}\Sigma^*_c$ and $\bar{D}^*\Sigma_c$ interactions and the LHCb hidden-charmed pentaquarks,''
  Phys.\ Lett.\ B {\bf 753}, 547 (2016)



\bibitem{Liu:2015fea}
  X.~H.~Liu, Q.~Wang and Q.~Zhao,
  ``Understanding the newly observed heavy pentaquark candidates,''
  Phys.\ Lett.\ B {\bf 757}, 231 (2016)


\bibitem{He:2016pfa}
  J.~He,
  ``Understanding spin parity of $P_c(4450)$ and $Y(4274)$ in a hadronic molecular state picture,''
  Phys.\ Rev.\ D {\bf 95}, 074004 (2017)

\bibitem{Wang:2011rga}
  W.~L.~Wang, F.~Huang, Z.~Y.~Zhang and B.~S.~Zou,
  ``$\Sigma_c \bar{D}$ and $\Lambda_c \bar{D}$ states in a chiral quark model,''
  Phys.\ Rev.\ C {\bf 84} (2011) 015203
  [arXiv:1101.0453 [nucl-th]].

  \bibitem{Yang:2011wz}
  Z.~C.~Yang, Z.~F.~Sun, J.~He, X.~Liu and S.~L.~Zhu,
  ``The possible hidden-charm molecular baryons composed of anti-charmed meson and charmed baryon,''
  Chin.\ Phys.\ C {\bf 36}, 6 (2012)
  [arXiv:1105.2901 [hep-ph]].  

\bibitem{Wu:2012md}
  J.~J.~Wu, T.-S.~H.~Lee and B.~S.~Zou,
  ``Nucleon Resonances with Hidden Charm in Coupled-Channel Models,''
  Phys.\ Rev.\ C {\bf 85} (2012) 044002
  [arXiv:1202.1036 [nucl-th]].



\bibitem{Wu:2010jy} J.~J.~Wu, R.~Molina, E.~Oset and B.~S.~Zou,
\textquotedblleft Prediction of narrow $N^{\ast }$ and $\Lambda ^{\ast }$
resonances with hidden charm above 4 GeV,\textquotedblright\ Phys.\ Rev.\
Lett.\ \textbf{105}, 232001 (2010).

\bibitem{Wang:2015qia} X.~Y.~Wang and X.~R.~Chen, ``The production of hidden
charm baryon $\text{N}^{\ast}(4261)$ from $\pi^{-}\text{p} \rightarrow\eta_%
\text{c}\text{n}$ reaction,'' EPL \textbf{109}, 41001 (2015).

\bibitem{Wang:2015xwa} X.~Y.~Wang and X.~R.~Chen, \textquotedblleft
Production of the superheavy baryon $\Lambda _{c\bar{c}}^{\ast }$ (4209) in
kaon-induced reaction,\textquotedblright\ Eur.\ Phys.\ J.\ A \textbf{51}, 85
(2015).

\bibitem{Huang:2013mua} Y.~Huang, J.~He, H.~F.~Zhang and X.~R.~Chen,
\textquotedblleft Discovery potential of hidden charm baryon resonances via
photoproduction,\textquotedblright\ J.\ Phys.\ G \textbf{41}, no. 11, 115004
(2014).

\bibitem{Kim:2016cxr} S.~H.~Kim, H.~C.~Kim and A.~Hosaka, \textquotedblleft
Heavy pentaquark states $P_{c}(4380)$ and $P_{c}(4450)$ in the $J/\psi $
production induced by pion beams off the nucleon,\textquotedblright\ Phys.\
Lett.\ B \textbf{763}, 358 (2016).

\bibitem{Lu:2015fva} Q.~F.~L\"{u}, X.~Y.~Wang, J.~J.~Xie, X.~R.~Chen and
Y.~B.~Dong, \textquotedblleft Neutral hidden charm pentaquark states $%
P_{c}^{0}(4380)$ and $P_{c}^{0}(4450)$ in $\pi ^{-}p\rightarrow J/\psi n$
reaction,\textquotedblright\ Phys.\ Rev.\ D \textbf{93}, 034009
(2016).

\bibitem{Garzon:2015zva} E.~J.~Garzon and J.~J.~Xie, \textquotedblleft
Effects of a N$_{c\overline{c}}^{\ast }$ resonance with hidden charm in the $%
\pi ^{-}p\rightarrow D^{-}\Sigma _{c}^{+}$ reaction near
threshold,\textquotedblright\ Phys.\ Rev.\ C \textbf{92}, 035201
(2015).

\bibitem{Karliner:2015voa} M.~Karliner and J.~L.~Rosner, \textquotedblleft
Photoproduction of Exotic Baryon Resonances,\textquotedblright\ Phys.\
Lett.\ B \textbf{752}, 329 (2016).

\bibitem{Wang:2015jsa} Q.~Wang, X.~H.~Liu and Q.~Zhao, \textquotedblleft
Photoproduction of hidden charm pentaquark states $P_{c}^{+}(4380)$ and $%
P_{c}^{+}(4450)$,\textquotedblright\ Phys.\ Rev.\ D \textbf{92}, 034022
(2015).

\bibitem{Meziani:2016lhg} Z.~E.~Meziani \textit{et al.}, \textquotedblleft A
Search for the LHCb Charmed `Pentaquark' using Photo-Production of $J/{\psi }
$ at Threshold in Hall C at Jefferson Lab,\textquotedblright\
arXiv:1609.00676 [hep-ex].

\bibitem{Wang:2019krd} X.~Y.~Wang, X.~R.~Chen and J.~He, \textquotedblleft
Possibility to study pentaquark states $P_{c}(4312),P_{c}(4440)$ and $%
P_{c}(4457)$ in $\gamma p\rightarrow J/\psi p$ reaction\textquotedblright\
arXiv:1904.11706 [hep-ph].

\bibitem{Jenkins:1977xb} K.~Jenkins \textit{et al.}, \textquotedblleft A
Search for the Reaction $\pi ^{-}p\rightarrow J/\psi n$ Near
Threshold,\textquotedblright\ Phys.\ Rev.\ D \textbf{17}, 52 (1978).

\bibitem{Chiang:1986gn} I.~H.~Chiang \textit{et al.}, \textquotedblleft
Search For Exclusiv $J/\psi $ Production,\textquotedblright\ Phys.\ Rev.\ D
\textbf{34}, 1619 (1986).

\bibitem{Austregesilo:2018mno} A.~Austregesilo [GlueX Collaboration],
``Light-Meson Spectroscopy at GlueX,'' Int.\ J.\ Mod.\ Phys.\ Conf.\ Ser.\
\textbf{46}, 1860029 (2018).

\bibitem{Kumano:2015gna} S.~Kumano, ``Spin Physics at J-PARC,'' Int.\ J.\
Mod.\ Phys.\ Conf.\ Ser.\ \textbf{40}, 1660009 (2016).

\bibitem{Nerling:2012er} F.~Nerling [COMPASS Collaboration],
\textquotedblleft Hadron Spectroscopy with COMPASS: Newest
Results,\textquotedblright\ EPJ Web Conf.\ \textbf{37} (2012) 01016.

\bibitem{Kim:2011rm} S.~H.~Kim, S.~i.~Nam, Y.~Oh and H.~C.~Kim,
\textquotedblleft Contribution of higher nucleon resonances to $K^{\ast }{%
\Lambda }$ photoproduction,\textquotedblright\ Phys.\ Rev.\ D \textbf{84},
114023 (2011).

\bibitem{Oh:2007jd} Y.~Oh, C.~M.~Ko and K.~Nakayama, \textquotedblleft
Nucleon and Delta resonances in K Sigma(1385) photoproduction from
nucleons,\textquotedblright\ Phys.\ Rev.\ C \textbf{77}, 045204 (2008).

\bibitem{Zou:2002yy} B.~S.~Zou and F.~Hussain, ``Covariant L-S scheme for
the effective N*NM couplings,'' Phys.\ Rev.\ C \textbf{67}, 015204 (2003).

\bibitem{Lin:1999ve} Z.~W.~Lin, C.~M.~Ko and B.~Zhang, ``Hadronic scattering
of charm mesons,'' Phys.\ Rev.\ C \textbf{61}, 024904 (2000).

\bibitem{Baru:2011bw} V.~Baru, C.~Hanhart, M.~Hoferichter, B.~Kubis,
A.~Nogga and D.~R.~Phillips, \textquotedblleft Precision calculation of
threshold $\pi ^{-}d$ scattering, pi N scattering lengths, and the GMO sum
rule,\textquotedblright\ Nucl.\ Phys.\ A \textbf{872}, 69 (2011).

\bibitem{Wu:2013xma} J.~J.~Wu and T.-S.~H.~Lee, \textquotedblleft Production
of $J/\psi $ on the nucleon and on deuteron targets,\textquotedblright\
Phys.\ Rev.\ C \textbf{88}, 015205 (2013).

\bibitem{Wang:2017qcw} X.~Y.~Wang and J.~He, \textquotedblleft Investigation
of pion-induced $f_{1}(1285)$ production off a nucleon target within an
interpolating Reggeized approach,\textquotedblright\ Phys.\ Rev.\ D \textbf{%
96}, 034017 (2017).

\bibitem{Liu:2008qx} X.~H.~Liu, Q.~Zhao and F.~E.~Close, \textquotedblleft
Search for tetraquark candidate Z(4430) in meson
photoproduction,\textquotedblright\ Phys.\ Rev.\ D \textbf{77}, 094005
(2008).

\bibitem{Wan:2015gsl} X.~Y.~Wang and J.~He, \textquotedblleft $K^{\ast
0}\Lambda $ photoproduction off a neutron,\textquotedblright\ Phys.\ Rev.\ C
\textbf{93}, 035202 (2016).

\bibitem{Wang:2015hfm} X.~Y.~Wang, J.~He and H.~Haberzettl,
\textquotedblleft Analysis of recent CLAS data on $\Sigma ^{\ast }(1385)$
photoproduction off a neutron target,\textquotedblright\ Phys.\ Rev.\ C
\textbf{93}, 045204 (2016).

\bibitem{Wang:2017plf} X.~Y.~Wang and J.~He, \textquotedblleft Analysis of
recent CLAS data on $f_{1}(1285)$ photoproduction,\textquotedblright\ Phys.\
Rev.\ D \textbf{95}, 094005 (2017).

\bibitem{Wang:2018mjz} X.~Y.~Wang, J.~He, Q.~Wang and H.~Xu,
\textquotedblleft Productions of $f_{1}(1420)$ in pion and kaon induced
reactions,\textquotedblright\ Phys.\ Rev.\ D \textbf{99}, 014020
(2019).

\bibitem{Haberzettl:2015exa} H.~Haberzettl, X.~Y.~Wang and J.~He,
\textquotedblleft Preserving Local Gauge Invariance with t-Channel Regge
Exchange,\textquotedblright\ Phys.\ Rev.\ C \textbf{92}, 055503 (2015).

\bibitem{Ozaki:2009wp} S.~Ozaki, H.~Nagahiro and A.~Hosaka, ``Charged K*
Photoproduction in a Regge model,'' Phys.\ Rev.\ C \textbf{81}, 035206
(2010).
\end{thebibliography}
\end{document}